\documentclass[proof]{pasj00}
\Received{\textit{2007 November 24}}
\Accepted{\textit{2008 April 10}}
\Published{$\langle$publication date$\rangle$}
\SetRunningHead{Yamamoto et al.}{CO clouds associated with jets}

\usepackage{times}

\begin{document}

\title{Aligned Molecular Clouds towards SS433 and L=348.5 degrees; \\
Possible Evidence for Galactic ''Vapor Trail'' Created by 
Relativistic Jet}
\author{Hiroaki \textsc{Yamamoto},\altaffilmark{1}
          Shingo \textsc{Ito},\altaffilmark{1}
          Shinji \textsc{Ishigami},\altaffilmark{1}
          Motosuji \textsc{Fujishita},\altaffilmark{1}
          Tokuichi \textsc{Kawase},\altaffilmark{1}
          Akiko \textsc{Kawamura},\altaffilmark{1}
          Norikazu \textsc{Mizuno},\altaffilmark{1}
          Toshikazu \textsc{Onishi},\altaffilmark{1}
          Akira \textsc{Mizuno},\altaffilmark{2}
          Naomi M. \textsc{McClure-Griffiths},\altaffilmark{3}
          and 
          Yasuo \textsc{Fukui}\altaffilmark{1}
}
\altaffiltext{1}{Department of Astrophysics, Nagoya University, 
                 Chikusa-ku, Nagoya, Aichi, 464-8602}
\altaffiltext{2}{Solar-Terrestrial Environment Laboratory, Nagoya 
                 University, Chikusa-ku, Nagoya, Aichi, 464-8601}
\altaffiltext{3}{Australia Telescope National Facility, CSIRO, P.O. 
                 Box 76, Epping, NSW 1710, Australia}

\email{hiro@a.phys.nagoya-u.ac.jp}

\maketitle

\begin{abstract}
We have carried out a detailed analysis of the NANTEN $^{12}$CO 
($J$=1--0) dataset at 4-arcmin resolution in two large areas of 
$\sim$ 25 square degrees towards SS433 ($l$ $\sim$ \timeform{40D}) 
and of $\sim$ 18 square degrees towards $l$ $\sim$ \timeform{348D.5}, 
respectively.  We have discovered two groups of remarkably aligned 
molecular clouds at high galactic latitudes of $\mid$$b$$\mid$ $\sim$ 
\timeform{1D}--\timeform{5D} in the two regions. In SS433, we have 
detected 10 clouds in total, which are well aligned nearly along the 
axis of the X-ray jet emanating from SS433. These clouds have similar 
line-of-sight velocities of 42 -- 56 km s$^{-1}$ in $V_{\rm LSR}$ and 
the total projected length of the feature is $\sim$ 300 pc, three 
times larger than that of the X-ray jet, at a distance of 3 kpc. 
Towards $l$ $\sim$ \timeform{348D.5}, we have detected four clouds 
named as MJG348.5 at line-of-sight velocities of $-$80 -- $-$95 
km s$^{-1}$ in $V_{\rm LSR}$, which also show alignment nearly 
perpendicular to the Galactic plane. The total length of the feature 
is $\sim$ 400 pc at a kinematic distance of 6 kpc. In the both cases, 
the CO clouds are distributed at high galactic latitudes, 
$\mid$$b$$\mid$ $\sim$ \timeform{1D}--\timeform{5D}, where such 
clouds are very rare. In addition, their alignments and coincidence 
in velocity should be even rarer, suggesting that they are physically 
associated. We tested a few possibilities to explain these clouds, 
including protostellar outflows, supershells, and interactions with 
energetic jets. Among them, a favorable scenario is that the 
interaction between relativistic jet and the interstellar medium 
induced the formation of molecular clouds over the last $\sim$ 
10$^{5-6}$ yrs. It is suggested that the timescale of the relativistic 
jet may be considerably larger, in the order of 10$^{5-6}$ yrs, than 
previously thought in SS433. The driving engine of the jet is 
obviously SS433 itself in SS433, although the engine is not yet 
identified in MJG348.5 among possible several candidates detected 
in the X-rays and TeV gamma rays.  

\end{abstract}

\section{Introduction}
Astrophysical jets of various scales are recognized as ubiquitous 
phenomena in the Universe and the physics of jets is one of the most 
fundamental issues in astrophysics.  On stellar scales, molecular jets 
driven by young protostars or bipolar outflows are well known 
phenomena since 1980's and are believed to represent a crucial step 
in the proto-stellar evolution (e.g., Lada 1985; Fukui 1989; Bally et 
al. 2005).  In addition, evolved compact stellar remnants like pulsars 
such as Crab pulsar, Vela pulsar, and MSH 15$-$52 exhibit more 
energetic relativistic jet with non-thermal radiation observed as 
pulsar-driven nebulae (e.g., Tamura et al. 1996; 
Weisskopf et al. 2000; Pavlov et al. 2003).  Even more massive 
objects, black-hole candidates such as SS433 and GRS1915+105, exhibit 
much more energetic relativistic jet whose velocity is close to the 
light speed (e.g., Margon 1984; Mirabel \& Rodr\'{i}guez 1994). 
Magneto-hydrodynamical numerical simulations on these jets have 
been carried out by several authors mainly in order to reproduce 
jets themselves (e.g., Kato, Hayashi \& Matsumoto 2004; 
Uzdensky \& MacFadyen 2006). Observationally, these relativistic 
jets have been detected so far only through high energy phenomena 
including non-thermal radio emission and X-rays but their interaction 
with the interstellar medium were very poorly known with a possible 
few examples in the Galaxy (e.g., Mirabel \& Rodr\'{i}guez 1999); 
we note that interactions with the much more diffuse gas are 
discussed in some of the jets in external galaxies (e.g., 
Oosterloo \& Morganti 2005; Krause et al. 2007). 

Some of the relativistic jets whose origin is a neutron star or a 
black hole have been identified (e.g., Mirabel et al. 1992; 
Rodr\'{i}guez, Mirabel \& Mart\'{i} 1992). A superluminal source, 
GRS1915+105, is a candidate for the relativistic jet interacting 
with the interstellar matter; Chaty et al. (2001) show that two 
IRAS point sources associated are located symmetrically at $\sim$ 60 
pc on the both sides of GRS1915+105 in a straight line. It is argued 
that the relativistic jet from GRS1915+105 whose velocity is 
estimated to be 0.92$c$ may have interacted with the molecular clouds 
and possibly induced star formation.

SS433 is an X-ray binary system consisting of a black hole candidate 
or a neutron star and is accelerating relativistic jet whose speed 
is 0.26$c$  (Margon \& Anderson 1989). SS433 is also associated with 
a symmetric and linear jet as observed in the X-ray. The jet is 
extended by about \timeform{1D} on each side of SS433, and the two 
lobes are called as East lobe and West lobe, respectively 
(Kotani 1997). The length of the jet is estimated to be about 40 pc 
on each side if we adopt a distance of $\sim$ 3 kpc 
(Dubner et al. 1998). It is also known that SS 433 is located towards 
the center of a supernova remnant (=SNR) W50. This SNR shows 
a barrel-type shape in non-thermal radio continuum emission. It has 
features called as "ears" and "wing" along the jet axis 
(Elston \& Baum 1987), which may be due to the interaction of the 
SNR with the surroundings.

The interaction of the SS433-W50 system and the surroundings has been 
studied by several authors. Dubner et al. (1998) studied the radio 
continuum emission at 1.4 GHz and the 21 cm H\emissiontype{I} 
emission, and noted that the H\emissiontype{I} at $V_{\rm LSR}$ 
$\sim$ 42 km s$^{-1}$ has a cavity-like shape surrounding the radio 
continuum emission, which may indicate some interaction between 
the H\emissiontype{I} and the SNR. On the other hand, Lockman et al. (2007)
suggests that SS433-W50 interacts with the H\emissiontype{I} at
$V_{\rm LSR}$ $\sim$ 75 km s$^{-1}$. Safi-Harb \& Ogelman (1997) and 
Safi-Harb \& Petre (1999) analyzed the X-ray data of ROSAT, ASCA, 
and RXTE, and discuss the interaction between the jet from SS 433 
and the surroundings. Band (1987) and Band \& Gordon (1989) studied 
the far infrared data taken with the IRAS and discovered some knots which may be associated with 
the jet, providing another piece of evidence for the interaction 
(Wang et al. 1990). Moldowan et al. (2005) suggest that one of 
the infrared objects is not correlated with the X-ray emission by 
the Chandra observation.

SS433 and its relationship with molecular clouds were studied by 
Huang et al. (1983). They claimed that the CO molecular clouds 
at $V_{\rm LSR}$ $\sim$ 27--36 km s$^{-1}$ may be spatially 
correlated with W50. On the other hand, Band et al. (1987) note 
that these clouds are located on the near side of and is not 
physically related with SS433. Subsequently, 
Durouchoux et al. (2000) suggest that CO clouds at 
$\sim$ 50 km s$^{-1}$ located towards the overlapped region of 
the West lobe of the X-ray jet may be interacting with the jet, 
because the X-ray hot spots are associated with them. 
Fuchs et al. (2002) observed the region with ISO and the IRAM 30m 
telescope to find that some of the IRAS knots are associated with 
these molecular clouds. One of these knots coincides with that 
noted by Durouchoux et al. (2000). Lockman et al. (2007) also studied
the association between the SS433-W50 and CO clouds using FCRAO 14m telescope
at 2 degrees $\times$ 2 degrees centered on SS433. They concluded that
there are no association between the SS433-W50 and the CO clouds.

To summarize, the previous works on the possible interaction 
between relativistic jet and the interstellar matter have shown 
the two possible cases of such interaction, SS433 and GRS1915+105. 
It is therefore still the beginning of observations of such 
interactions and theoretical studies of the 
interactions largely remain unexplored. 

NANTEN, a 4m mm/sub-mm telescope located in Chile, has been used 
to make an extensive survey of the Galactic plane in 
$^{12}$CO ($J$=1--0) emission at a grid spacing of \timeform{4'} 
at $\mid$$b$$\mid$ $\leqq$ \timeform{5D} and that of \timeform{8'} 
at \timeform{5D} $\leqq$ $\mid$$b$$\mid$ $\leqq$ \timeform{10D} 
(for more details see e.g., Mizuno \& Fukui 2004). This offers a 
few times spatially finer CO images at 
$\mid$$b$$\mid$ $\leqq$ \timeform{5D} compared to the previous 
low resolution CO survey (Dame, Hartmann \& Thaddues 2001).  
The new NANTEN CO dataset is useful to search uniformly for various 
phenomena towards $\pm$\timeform{10D} of the Galactic plane on 
a large scale including supershells and other active events 
(e.g., Fukui et al. 1999, 2006; Matsunaga et al. 2001). 

In the present paper, we show and discuss the observational 
results of molecular clouds which may be associated with relativistic 
jet toward SS433 and $l$ $\sim$ \timeform{348D.5} , and their 
implications. Section 2 gives a summary of the CO dataset. Sections 3 
and 4 present the main observational results for the two regions and 
a model is presented and discussed in section 5. Conclusions are given 
in section 6.

\section{The NANTEN CO Dataset}

We used the NANTEN Galactic Plane Survey dataset of the 
$^{12}$CO ($J$=1--0) emission (Mizuno \& Fukui 2004). 
The coverage of the data is 220 degrees in $l$ from $l$ $\sim$ 
\timeform{200D} to \timeform{60D} including the Galactic center 
and at $\mid$$b$$\mid$ $\leqq$ \timeform{5D} with a grid spacing 
of \timeform{4'} and at \timeform{5D} $\leqq$ $\mid$$b$$\mid$ $\leqq$ 
\timeform{10D} with a grid spacing of \timeform{8'} for the main beam 
width of \timeform{2'.6} in addition to the selected areas of nearby 
clouds at $\mid$$b$$\mid$ $\geqq$ \timeform{10D} 
(e.g., Mizuno et al. 2001; Onishi et al. 2001).  The total number of 
observed points is $\sim$ 1.1 million.  The velocity coverage and 
resolution of the data are usually from 300 km s$^{-1}$ to $-$300 km s$^{-1}$ and 0.65 
km s$^{-1}$, respectively.  All the observations were carried out by 
the position switching technique. The telescope was equipped with a 
superconducting mixer receiver (Ogawa et al 1990).  The system 
temperature including the atmosphere was in a range of 250--350 K in 
the single side band mode (=SSB) on average towards the zenith and 
the typical r.m.s. noise fluctuations of the spectral data are $\sim$ 
0.35 K/(0.65 km s$^{-1}$) in the absolute antenna temperature, 
$T_{\mathrm{R}}^{*}$, corresponding to an integration time of a few 
to several seconds per point at on-position.

\section{SS433}
\subsection{Large Scale Distribution of Molecular Clouds}

Figure 1 shows the $^{12}$CO($J$=1--0) integrated intensity 
distribution whose velocity range is $V_{\rm LSR}$ = 40 to 60 
km s$^{-1}$ in the region of $l$ = \timeform{37D} to \timeform{42D} 
and $b$ = \timeform{-5D} to \timeform{0D}. Superposed are the 
H\emissiontype{I} integrated intensity distribution at a 16 arcmin 
effective resolution with the Parkes 64m telescope 
(McClure-Griffiths et al. 2005) and ASCA X-ray distribution of 
the GIS image in the 0.7--10 keV band (Kotani 1998). We have 
identified ten $^{12}$CO clouds along the X-ray jet axis of SS433 
at the lowest 3$\sigma$ contour level in 
$l$ $\sim$ \timeform{39D} -- \timeform{41D.5} and 
$b$ $\sim$ \timeform{-5D} -- \timeform{-1D}.  In the southeast of 
SS433 there are six clouds as named from SS433-S1 to SS433-S6 at 
$b$ $\sim$ \timeform{-3D} -- \timeform{-5D} and at $V_{\rm LSR}$ 
$\sim$ 42 -- 45 km s$^{-1}$. In the northwest of SS433 there are 
four clouds as named from SS433-N1 to SS433-N4 at 
$b$ $\sim$ \timeform{-1D} -- \timeform{-2D} and at 
$V_{\rm LSR}$ = 50 -- 55 km s$^{-1}$. The observed parameters of 
these clouds are listed in Table 1. 

These ten clouds exhibit a remarkably straight distribution along 
the axis of the X-ray jet of SS433. This alignment is approximated 
by a dashed line in Figure 1 as determined by a linear regression 
fit on the clouds weighted in the total CO intensity and is expressed 
as $b$(\timeform{D}) = (63$\pm$2)(\timeform{D}) $-$ 
(1.65$\pm$0.05)$\times$$l$(\timeform{D}) with a high correlation 
coefficient of $\sim$ 0.98. This line passes almost exactly through 
the position of SS433 
($l$, $b$)=(\timeform{39D.7}, $-$\timeform{2D.2}) as shown in 
Figure 1. We note that the position angle of the line is 
$\sim$ \timeform{30D} in the galactic coordinate, while that of the 
X-ray jet is $\sim$ \timeform{20D}, showing a small difference of 
$\sim$ \timeform{10D}. 

The southern clouds have been discovered by the present work. 
Some of the northern clouds have already been observed and the 
interaction with the SS433 jet has been discussed for SS433-N1 
to SS433-N4 by the previous authors (Band et al. 1989; Durouchoux 
et al. 2000; Fuchs et al. 2002; Chaty et al. 2001; 
Moldowan et al. 2005). 

Figure 2(b) shows the velocity distribution of the $^{12}$CO clouds 
in a position-velocity diagram superposed on the H\emissiontype{I} 
distribution. The position is the offset from SS433 taken along a 
line tilted to the galactic plane by \timeform{45D} as indicated 
in Figure 2(a). Some of the clouds are overlapped in the figure and 
the number of the clouds apparently becomes less than ten. The typical 
velocities of the northern clouds, SS433-N1 to SS433-N4, and the 
southern clouds, SS433-S1 to SS433-S6, are $\sim$ 53 km s$^{-1}$ and 
$\sim$ 43 km s$^{-1}$, respectively, on average. We note that the 
internal velocity dispersions of these clouds are as small as 
2 km s$^{-1}$ while that of SS433-S6 is $\sim$ 4 km s$^{-1}$. It is 
not certain if another cloud at the offset of \timeform{1D.5} and 
$V_{\rm LSR}$ $\sim$ 50 km s$^{-1}$ is related to the present 
northern clouds.

Figure 3 shows a larger scale view of the region in the $^{12}$CO 
integrated intensity distribution in the same velocity range as in 
Figure 1, covering 50 square degrees from $l$ = \timeform{35D} to 
\timeform{45D} and $b$ = \timeform{-5D} to \timeform{0D}.  This 
demonstrates that there are only a few CO clouds except for the 
present ones over a volume of $\sim$ 500 pc (in $l$) $\times$ 
$\sim$ 100 pc (in $b$) $\times$ $\sim$ 1 kpc in the line of sight 
(in $v$) at $b$ $\lesssim$ \timeform{-3D} for an assumed distance 
of $\sim$ 3 kpc (see section 3.2). We note that the present CO 
southern clouds are distributed up to $z$ $\sim$ 240 pc, very far 
out from the galactic plane whose typical scale height in the CO 
emission is $\sim$ 87 pc (Dame et al. 1987). The southern CO clouds 
located at around \timeform{-4D} are therefore very rare and a group 
of such aligned clouds is quite unique.  We shall hereafter assume 
that the ten clouds are at the distance of SS433 because of their 
similar velocities and alignment.

\subsection{Physical parameters}

The two average velocities of the CO clouds, 53 km s$^{-1}$ and 
43 km s$^{-1}$, correspond to kinematic distances of 3.5 kpc and 
3 kpc, respectively, for the flat rotation curve (Brand \& Blitz 1993) 
while another kinematic distance around $\sim$ 10 kpc is also 
permitted.  According to the previous studies, the distance to SS433 
is estimated to be $\sim$ 3 kpc from the absorption of atomic 
hydrogen (Gorkom, Goss, Shaver 1980) and the velocity of the cavity 
of atomic hydrogen (Dubner et al. 1998), $\sim$ 4.85 kpc from six 
radio continuum images at two day intervals with VLBI by 
Vermeulen et al. (1993) and  $\sim$ 5.5 kpc from the radio continuum 
observations in the central part of SS433 with VLA 
(Hjellming \& Johnston 1980, Blundell \& Bowler 2004).  We shall 
hereafter tentatively adopt the distance to SS433 the smaller value, 
3 kpc, which is consistent with the present kinematic distance.  
This gives conservative estimates of related physical parameters such 
as the radius and mass of the molecular clouds. If the cloud velocity 
is affected by motion other than the galactic rotation, this estimate 
needs to be reconsidered. 

The physical parameters of the clouds are calculated as listed in 
Table 2. We adopt the X factor, which is defined as 
$N$(H$_{\rm 2}$)/$W$($^{12}$CO), of 
2.0$\times$10$^{20}$ cm$^{-2}$/(K km s$^{-1}$) (Lebrum et al. 1983; 
Bertsch et al. 1993), in these calculations.  The linewidth, mass, 
size, and peak $T_{\mathrm{R}}^{*}$ are similar to those of nearby 
dark clouds (e.g., Mizuno et al. 2001; Tachihara et al. 2001). 
The cloud mass ranges from $\sim$ 10$^2$ to $\sim$ 10$^3$ $M_{\odot}$. 
The total molecular mass of the southern clouds amounts to 
$\sim$ 5.9$\times$10$^3$ $M_{\odot}$ while that of the northern clouds 
$\sim$ 9.0$\times$10$^3$ $M_{\odot}$. The virial mass of these 
molecular clouds is several times larger than this luminous mass, 
indicating that the molecular clouds are not gravitationally bound; 
this is a typical property of molecular clouds in such a low mass 
range (e.g., Onishi et al. 2001; Yamamoto et al. 2003, 2006).  
They may be confined by the ambient pressure or be transient at a 
timescale of $\sim$ 10$^6$ yrs, the crossing timescale as discussed 
in the previous works including the above two.

\subsection{Comparison with radio continuum, H\emissiontype{I} 
            and X-ray around SS433}

Figure 4 shows an overlay of the radio continuum distribution of W50, 
a supernova remnant associated with SS433, at 4850 MHz 
(e.g., Condon et al. 1989) on the H\emissiontype{I} 21 cm line 
integrated intensity in the $V_{\rm LSR}$ range 40 -- 60 km s$^{-1}$. 
The SNR is elongated by $\sim$ \timeform{2D} towards the same direction 
as the X-ray jet while the width is $\sim$ \timeform{1D}, 
significantly larger than the X ray, showing a sharp intensity 
gradient towards the galactic plane. The northern clouds SS433-N1, 
SS433-N2 and SS433-N3 are located towards the northern edge of 
the SNR and SS433-N4 seems to be located inside of the SNR.  
On the other hand, the southern clouds are separated from the SNR. 
Dubner et al. (1998) show that the SNR W50 is located towards a hole 
of the H\emissiontype{I} emission and argue that the hole may have 
been created by the supernova explosion. We confirm this suggestion 
as the H\emissiontype{I} depression towards SS433 in Figure 4. Recent
H\emissiontype{I} study in and around the SS433-W50 reported by Lockman et al. (2007)
suggests that the H\emissiontype{I} gas at a velocity $\sim$ 75 km s$^{-1}$ 
has interacted with the W50 from the morphology of the H\emissiontype{I} gas by new observations and its distance.
The 75 km s$^{-1}$ component of the H\emissiontype{I} gas also looks like interacting with SS433-W50 at upper part
of the W50. But here we use the H\emissiontype{I} data at $V_{\rm LSR}$ $\sim$ 40--60 km s$^{-1}$ because the
velocity of present molecular clouds is $\sim$ 40--60 km s$^{-1}$ and these are not associated with the H\emissiontype{I}
at $V_{\rm LSR}$ $\sim$ 75 km s$^{-1}$.

We find in Figures 1 and 3 that the southern CO clouds are located 
towards a H\emissiontype{I} protrusion peaked at ($l$, $b$) $\sim$ 
(\timeform{40D.9}, \timeform{-3D.9}), where the present CO clouds 
are distributed within a H\emissiontype{I} contour of 410 
K km s$^{-1}$. Figure 2(b) shows that the CO velocity agrees with 
the H\emissiontype{I} velocity at the offset of \timeform{-1D} -- 
\timeform{-3D}. These suggest that the southern CO clouds and 
the H\emissiontype{I} are physically associated. In Figure 2(b) 
the northern clouds also seem to be associated with 
the H\emissiontype{I} at the offset of \timeform{0D.5} -- 
\timeform{1D}, while the HI velocity is slightly larger by a few 
km s$^{-1}$ than the CO velocity.

The H\emissiontype{I} mass of the protrusion towards \timeform{39D} 
$\leqq$ $l$ $\leqq$ \timeform{41D.4} and 
\timeform{-3D.4} $\leqq$ $b$ $\leqq$ \timeform{-5D} within a contour 
level of 410 K km s$^{-1}$ is estimated to be 51000 $M_{\odot}$ by 
using the conventional factor of the HI intensity into mass, 
1.8$\times$10$^{18}$ cm$^{-2}$/(K km s$^{-1}$).  The total molecular 
mass of the southern CO clouds is $\sim$ 5900 $M_{\odot}$, 
corresponding to about one tenth of that of H\emissiontype{I}. 

ASCA X-ray distribution shown in Figure 1 indicates clearly the two 
lobes of the X-ray jet, the West and East lobes, respectively. 
SS433-N4 is located towards the West lobe and SS433-N1 and SS433-N3 
show some overlapping with this lobe also. This may suggest some 
interaction between them as already noted by 
Durouchoux el al. (2000).  SS433-N2 is somewhat far from the other 
clouds and its physical association with the X-ray lobe may not be 
certain.  We note that the field of view of the X-ray observations 
towards the lobe is limited to the three fields centered on SS433 
covering regions within $\sim$ \timeform{1D} of SS433.  So, the 
distribution of the X-ray outside the area remains to be uncovered.

\section{Linearly Aligned Molecular Clouds Perpendicular to the 
         Galactic Plane towards $l$ $\sim$ \timeform{348D.5}}

\subsection{Large Scale Molecular Distribution}

In the course of a detailed analysis of the $^{12}$CO dataset, 
we have discovered an unusual aligned distribution of CO clouds 
toward $l$ $\sim$ \timeform{348D.5}.  Figure 5 shows an integrated 
intensity distribution of $^{12}$CO ($J$=1--0) emission whose velocity 
range is from $V_{\rm LSR}$ = $-$100 to $-$70 km s$^{-1}$ in the 
region of \timeform{347D} $\leqq$ $l$ $\leqq$ \timeform{350D} and 
\timeform{-3D} $\leqq$ $b$ $\leqq$ \timeform{3D}.  Four clumpy clouds 
located at $b$ $\sim$ \timeform{1D.7}, \timeform{-0D.8}, 
\timeform{-1D.7}, and \timeform{-2D.1} are aligned nearly 
perpendicular to the Galactic plane where most of the intense CO 
emission is confined at $\mid$$b$$\mid$ $\lesssim$ \timeform{0D.5}. 
We shall tentatively assume the physical association of the four 
clouds in the following and name the group of the southern and 
northern clouds as MJG348.5 
(=molecular jet at $l$ = \timeform{348D.5}). We call hereafter 
the four components, MJG348.5-N, MJG348.5-S1, MJG348.5-S2, and 
MJG348.5-S3, respectively, as labeled in Figure 5.  We note the 
separations between the northern and southern tips of the 
molecular clouds, MJG348.5-N and MJG348.5-S3, and the Galactic 
plane is nearly $\sim$ \timeform{2D}, respectively.  A linear 
regression fit to the four clouds, shown in Figure 5 by a dashed 
line, with integrated intensity weighting yields a relationship, 
$b$(\timeform{D})=($-$13.67$\pm$1.63)$\times$$l$(\timeform{D})+
(4765.69$\pm$566.64)(\timeform{D}) with a correlation coefficient 
of $\sim$ 0.89, which passes through the Galactic plane at 
$l$ = \timeform{348D.52} and is nearly perpendicular to the 
Galactic plane at an angle between the line and the Galactic 
plane of $\sim$ \timeform{86D}. The apparent largest separation 
of MJG348.5-S3 from the Galactic plane is $\sim$ \timeform{2D.4}. 
The $V_{\rm LSR}$ of the southern three components is $\sim$ $-$82 
km s$^{-1}$ and that of the northern one is $\sim$ $-$95 km s$^{-1}$ 
as shown in Figure 6(a).  We note that the two clouds, MJG348.5-N 
and MJG348.5-S3, exhibit the largest dispersion of $\sim$ 7 
km s$^{-1}$ (see Figure 6(b) and (c)), while the other clouds, 
MJG348.5-S1 and MJG348.5-S2, show smaller velocity dispersions of 
$\sim$ 3 km s$^{-1}$. 

We further note that another cloud is located at $l$ $\sim$ 
\timeform{347D.7} and $b$ $\sim$ \timeform{1D.6}  at a LSR velocity 
of $\sim$ $-$83.0 km s$^{-1}$. This cloud is also unusual at such high 
latitude and large LSR velocity. We come back to this feature later in 
comparison with the H\emissiontype{I} and in the discussion section.

Figure 7 shows a larger scale view of the region in the $^{12}$CO 
integrated intensity distribution in the same velocity range as in 
Figure 5, covering 200 square degrees from $l$ = \timeform{339D} to 
\timeform{359D} and $b$ = \timeform{-5D} to \timeform{5D}.  This 
demonstrates that there are only a few CO clouds except for the 
present ones over a volume of $\sim$ 2 kpc (in $l$) $\times$ $\sim$ 
0.4 kpc (in $b$) $\times$ $\sim$ 0.9 kpc in the line of sight (in $v$) 
at $\mid$$b$$\mid$ $\leqq$ \timeform{3D} for the kinematric distance 
of $\sim$ 6 kpc as derived below. The CO clouds located at more than 
\timeform{\pm1D.5} in Galactic latitude are very rare and a group of 
such clouds aligned in a straight line is unique as in case of SS433. 
We shall further note that the present CO clouds are distributed to 
$z$ $\sim$ 240 pc, very far out of the typical CO scale height of the 
CO emission, $\sim$ 87 pc (Dame et al. 1987).

\subsection{Physical Properties of the Molecular Clouds}

The averaged $V_{\rm LSR}$ of the MJG348.5-S1, MJG348.5-S2, and 
MJG348.5-S3 is $\sim$ $-$82 km s$^{-1}$, corresponding to a kinematic 
distance of 5.9 kpc or 10.8 kpc, and that of MJG348.5-N of 
$\sim$ $-$95 km s$^{-1}$ corresponds to 6.1 kpc or 10.5 kpc, 
respectively, if we assume the flat rotation curve 
(Brand \& Blitz 1993).  We shall tentatively adopt the smaller 
averaged value, $\sim$ 6 kpc, hereafter, since it gives conservative 
estimates of the cloud parameters, noting that the kinematic distance 
may include a large uncertainty toward this direction near the center. 
For instance, the kinematic distance changes from 5.7 kpc to 6.0 kpc 
for a velocity difference of +5 km s$^{-1}$, typical velocity 
dispersion in the H\emissiontype{I} clouds.  The alignment of the 
three southern molecular clouds is remarkably good over a length of 
$\sim$ 200 pc with a width of $\sim$ 10 pc at 6 kpc. The location of 
the northern cloud, MJG348.5-N, is fairly symmetric to MJG348.5-S3 with respect to 
the plane.

The relevant observed and derived physical parameters of the present 
clouds are summarized in Table 3. The peak temperature and line width 
of the $^{12}$CO emission of the four molecular clouds are 1.6 to 
3.3 K and 3.3 to 5.4 km s$^{-1}$, not much different from those of 
the typical CO clouds whose average density and kinetic temperature 
are $n$(H$_{\rm 2}$) $\sim$ 10$^{2}$ cm$^{-3}$ and 
$T_{\mathrm{k}}$ $\sim$ 10 K (see for typical nearby dark clouds, 
e.g., Mizuno et al. 2001; Tachihara et al. 2001). The range of mass 
of each CO cloud is from $\sim$ 1.5$\times$10$^{3}$ to 
$\sim$ 1.4$\times$10$^{4}$ $M_{\odot}$ by adopting X factor 
$N$(H$_2$)/$W$($^{12}$CO), of 
2.0$\times$10$^{20}$ cm$^{-2}$/(K km s$^{-1}$) (Bertsch et al. 1993). 
In total, the molecular mass in the four clouds is estimated to be 
$\sim$ 2.6$\times$10$^{4}$ $M_{\odot}$. They share similar dynamical 
properties with those in SS433 as mentioned in Section 3.2.

\subsection{Comparison with the H\emissiontype{I}}

Figures 8(a) and (b) show two velocity channel maps of $^{12}$CO 
superposed on the distribution of the 21 cm H\emissiontype{I} line 
emission integrated in the corresponding velocity ranges. The 
effective H\emissiontype{I} resolution is 16 arcmin with the Parkes 
64m telescope (McClure-Griffiths et al. 2005).  The strong 
H\emissiontype{I} emission at $\mid$$b$$\mid$ $\lesssim$ \timeform{1D} 
is the Galactic disk emission. At 
$\mid$$b$$\mid$ $\gtrsim$ \timeform{1D} we are able to identify 
the H\emissiontype{I} features associated with the present CO clouds. 
Towards MJG348.5-N at $b$ $\sim$ \timeform{1D.7}, an isolated 
H\emissiontype{I} cloud is found in Figure 8(a).  
This H\emissiontype{I} cloud having a size of $\sim$ 50 pc $\times$ 
$\sim$ 30 pc at 60 K km s$^{-1}$ is elongated in a similar direction 
to the CO distribution from NW to SE tilted to the Galactic plane by 
$\sim$ 45 degrees. Towards the southern three clouds, MJG348.5-S1, 
MJG348.5-S2, and MJG348.5-S3, we see that a protrusion of 
the H\emissiontype{I} emission is extended at $l$ $\sim$ 
\timeform{348D.5} up to $b$ $\sim$ \timeform{-2D.4}. MJG348.5-S2 is 
associated with part of the H\emissiontype{I} protrusion  towards 
$l$ $\sim$ \timeform{348D.6} and $b$ $\sim$ \timeform{-1D.7} at 
a contour level of $\sim$ 135 K km s$^{-1}$.  This H\emissiontype{I} 
emission has a size of $\sim$ 100 pc $\times$ $\sim$ 60 pc at 
the H\emissiontype{I} contour level of 105 K km s$^{-1}$ elongated 
to the Galactic plane by $\sim$ 45 degrees from NE to SW. 
The H\emissiontype{I} protrusion is also identified in Figure 6(a) 
in a velocity range from $-$90 km s$^{-1}$ to $-$75 km s$^{-1}$ at 
the H\emissiontype{I} contour level of $\sim$ 85 K degree. 
In addition, MJG348.5-S3 is associated with the southern extension 
of the H\emissiontype{I} protrusion towards $l$ $\sim$ 
\timeform{348D.6} and $b$ $\sim$ \timeform{-2D.0} (Figure 8(b)), 
which is identified at $V_{\rm LSR}$ $\sim$ $-$80 km s$^{-1}$ in 
Figure 6(a).  These spatial and velocity extents of 
the H\emissiontype{I} emission is to be regarded as the lower limits 
by considering the possible more extended H\emissiontype{I} features 
at lower intensity levels.  

We have further inspected the H\emissiontype{I} distribution 
in detail. Figure 9 shows the longitude-velocity diagrams of 
the H\emissiontype{I} superposed on the CO in $b$ from 
\timeform{2D.53} to \timeform{-2D.53} except for the area within 
$\pm$0.53 degrees where contamination is strong. The CO emission 
is clearly associated with H\emissiontype{I} at panels (d), (e), 
(l), (p), (q), (r), and (s) in Figure 9, confirming that the 
H\emissiontype{I} is associated with MJG348.5-S3. 
The H\emissiontype{I} associated with MJG348.5-S3 is not clearly 
resolved but the upper limits of the line width and the radius of 
the H\emissiontype{I} associated with MJG348.5-S3 are roughly 
estimated to be $\sim$ 5 km s$^{-1}$ and $\sim$ 6.8 arcmin = 12 pc 
at 6 kpc, respectively, yielding the crossing time of 
the H\emissiontype{I}, $\sim$ 2.3$\times$10$^{6}$ yrs, similar to 
that of the molecular cloud, MJG348.5-S3. The mass of 
the associated H\emissiontype{I} both in the southern protrusion 
and in the northern cloud is estimated to be $\gtrsim$ 25000 
$M_{\odot}$ at an H\emissiontype{I} integrated intensity level of 
100 K km s$^{-1}$, which is about two thirds of the total mass of 
the four molecular clouds. Here, we used the conventional relationship 
to convert the H\emissiontype{I} intensity into mass 
1.8$\times$10$^{18}$ cm$^{-2}$ (K km s$^{-1}$)$^{-1}$ by assuming 
optically thin H\emissiontype{I} emission. 

We shall note that the CO cloud towards $l$ $\sim$ 
\timeform{347D.7} and $b$ $\sim$ \timeform{1D.6} at a LSR velocity 
of $-$83.0 km s$^{-1}$ show H\emissiontype{I} counterpart. 
This cloud is clearly associated with H\emissiontype{I} as seen 
in panel (e) and (f) of Figure 9. Figure 8 also shows this 
H\emissiontype{I} cloud in panel (b). We further note that an 
H\emissiontype{I} protrusion is seen towards 
($l$, $b$) = (\timeform{347D.5}, \timeform{-1D.4}) of Figure 8 
which could be a possible counterpart in the south. The real 
physical association of these clouds are yet uncertain compared 
with MJG348.5 clouds.

To summarize, the H\emissiontype{I} distribution superposed in 
the position-velocity diagrams (Figures 6(a) and 9) and in the 
sky (Figure 8(a) and (b)) suggest that H\emissiontype{I} gas is 
physically associated with the four molecular clouds.

\subsection{High Energy Objects in the Galactic Plane}

Energetic sources are not associated towards the individual 
molecular clouds but toward the Galactic plane at 
$l$ $\sim$ \timeform{348D.5}, the point of crossover between 
the straight line defined by these molecular clouds (Section 4.1) 
and the Galactic plane, there are several energetic sources already 
confirmed by observations of non-thermal radio emission, X-rays and 
$\gamma$-rays as summarized in Table 4.  In Figure 5, the positions 
of the known supernova remnants, SNR G348.5+0.1 (CTB37A), 
G348.5$-$0.0, and G348.7+0.3 (CTB37B) are shown by crosses 
(e.g., Kassim et al. 1991). In addition, there are two 
TeV $\gamma$-ray sources confirmed by H.E.S.S. observations which 
may be associated with these SNRs or their stellar remnants as shown 
in Figure 15 of Aharonian et al. (2006).  We note that the positional 
coincidence of the crossover point and these energetic sources, 
in particular CTB37A, is remarkably good.  Other high energy 
sources are identified by EGRET and ROSAT All Sky Survey 
(Hartman et al. 1999; Voges et al. 1999) near the three SNRs 
although positions between these sources and the three SNRs are 
slightly different except for 1RXS J171354.4$-$381740 as shown 
in Figure 5 (see also Table 2). The relationship between molecular 
clouds and these energetic sources will be discussed in the next 
section.

\section{Discussion}

\subsection{The Aligned Molecular Clouds}

The present study has revealed remarkably well aligned molecular 
clouds of $\sim$ 300--400 pc in length in the two fields of 
the Galaxy. The existence of the CO clouds itself is rare at 
high $z$ greater than $\sim$ 100 pc. The alignments are not 
accidental by considering the even rare coincidence in velocity; 
the several clouds on a line passing through SS433 at 
$z$ $\sim$ 200 pc strongly indicate the physical association among 
them and also with SS433.  The similar features in MJG348.5 are also 
very unusual, suggesting their physical association.  The probability for
such aligned clouds is indeed very low as shown in Appendix.

We shall here estimate the spatial and velocity dispersions of 
the aligned molecular clouds.  In SS433, only the southern clouds 
are dealt with because the northern clouds are contaminated by 
the galactic background emission. The 1$\sigma$ dispersion of 
the displacement of the molecular clouds from the dashed line 
in Figure 1 is around $\sim$ 9.7$\pm$2.0 pc for the 
intensity-weighted average. Figure 2(b) was used to estimate 
1$\sigma$ velocity dispersion as $\sim$ 5.0$\pm$1.0 km s$^{-1}$. 
The same procedure was applied to the MJG348.5 clouds, yielding 
spatial dispersion of $\sim$ 7.3$\pm$1.4 pc and velocity dispersion 
of $\sim$ 2.2$\pm$0.3 km s$^{-1}$, respectively. These small 
dispersions in the both cases indicate the excellent alignments 
both in space and velocity (Table 1). 

SS433 is one of the best-studied compact objects driving 
relativistic jet. The jet of SS433 has a spatial extent of 
$\sim$ 80 pc in the X-ray, the largest size of such jet known 
to date in the Galaxy. The jet is still being accelerated near 
the driving source at a velocity of 0.26$c$ as determined from 
the Doppler measurements of H$\alpha$ emission (Margon et al. 1979). 
The jet is likely driven by the accretion disk plus a stellar remnant 
which has a deep gravitational potential well of a black hole or a 
neutron star.  The disk material is perhaps being supplied from a 
counterpart of the binary, an ordinary evolved star having a large 
envelope. On the other hand, MJG348.5 does not have a known 
jet-accelerating object in the center.  We shall first focus on 
SS433 in the following and discuss a possibility that the present 
clouds were created by the relativistic jet driven by a compact 
object and then we shall extend the model to MJG348.5. 

Before moving to the relativistic-jet interpretation, we shall 
consider two alternative possibilities, i.e., "protostellar 
bipolar outflow" and "supershell wall" to explain the aligned 
clouds. The known molecular outflow from young stars is several 
pc at most in length, nearly two orders of magnitude smaller 
than in the present two cases of SS433 and MJG348.5 
(e.g., Lada 1985; Fukui 1989; Bally et al. 2005), and exhibits 
the broad linewidths in the order of 10--100 km s$^{-1}$, more 
than an order of magnitude larger than the present linewidths. 
The aligned clouds are therefore quite different from the protostellar 
outflow. Supershells produced by massive stars via supernovae and/or 
stellar winds may offer an explanation for the high $z$ distribution 
and the large extents of $\sim$ 100 pc. The known molecular 
supershells are indeed characterized by a few 100 pc radius 
(Fukui et al. 1999; Yamaguchi et al. 1999; Matsunaga et al. 2001) 
but the straight distribution of the present clouds are hard to be 
reconciled with part of an expanding shell which should show 
non-uniform curved patterns in space and velocity typical to shells. 
We shall not discuss further on the supershell interpretation.

\subsection{The relativistic-jet model}

\subsubsection{Scenario}

The origin of the molecular clouds towards the SS433 region is 
explained as follows. Relativistic jet whose expanding velocity is 
a few tens \% of the light speed, part of which is observed as the 
X-ray jet,  interacted with the H\emissiontype{I} gas peaked towards 
($l$, $b$)= (\timeform{40D.9}, \timeform{-3D.9}).  The interaction 
with the jet agitated the pre-existent H\emissiontype{I} gas 
dynamically and heated it up significantly. The kinetic power of 
the SS433 jet $\sim$ 1.1$\times$10$^{46}$ erg/yr is in fact huge 
(Table 3); we shall tentatively assume that very hot gas such as 
observed in the X-ray jet is created via the interaction since any 
detailed calculations on the process are not found in the literature. 
A natural consequence of the interaction is a cylindrical expanding 
shock front compressing the gas, which leads to formation of the 
molecular clouds around the jet axis. The measured velocity and 
spatial dispersions of the clouds in Table 1 indicate that the 
typical expansion velocity and radius of the expanded cylinder 
are 2--5 km s$^{-1}$ and 7--10 pc, respectively. The timescale is 
then estimated roughly to be a few Myr by dividing the radius with 
the velocity.   This offers an explanation on the straight 
distribution of the present southern CO clouds. The spatial 
coincidence of the southern CO clouds with the H\emissiontype{I} 
protrusion over a length of $\sim$ 60 pc is consistent with this 
scenario because the background atomic gas is the necessary condition 
to form molecular clouds.  The observed clumped CO distribution may 
be due to the initial density inhomogeneities in H\emissiontype{I}, 
which is not yet resolved with the present H\emissiontype{I} beam in 
Figure 1, or due to the gravitational instability in the 
shock-compressed H\emissiontype{I} gas.  A similar process may have 
taken place in the north to form the northern clouds where the higher 
H\emissiontype{I} density near the galactic plane. The southern jet 
is extended at least $\sim$ 150 pc, while the northern jet can be 
traced up to $\sim$ 50 pc. This asymmetry may be ascribed to the 
increased deceleration near the galactic plane. 

The length of the observed X-ray jet indicates that the jet has a 
momentum large enough to travel over at least $\sim$ 40 pc. The 
present scenario implies that the relativistic jet of SS433 has an 
actual full extent of $\sim$ 150 pc on the southern side, 
significantly larger than the known size of the X-ray jet. 
We note that the X-ray observations by Kotani (1998) does not cover 
the area of the present southern clouds at $b$ less than 
\timeform{-2D} and the X-ray observations yet remain to be extended 
towards the region of the southern clouds. 

In order to explain the distribution of the aligned clouds toward 
$l$ $\sim$ \timeform{348D.5}, we argue that the same mechanism as 
in SS433 is working there by assuming the existence of relativistic 
jet and a compact driving engine similar to SS433. All the basic 
aspects of the interaction in SS433 are then applicable to MJG348.5. 
The background H\emissiontype{I} is also rich towards the CO clouds 
as is consistent with the model. The number of the clouds, four, 
in MJG348.5 is less than in SS433. This may be due to the lower 
spatial resolution at 6 kpc and/or due to the difference in the 
initial H\emissiontype{I} distribution.

\subsubsection{Timescales}

We may estimate the timescale of the interaction. An obvious one 
is the traveling time over $\sim$ 200 pc as given by 
$\gtrsim$ (150--240 pc)/0.26$c$ $\sim$ 3$\times$10$^{3}$ yrs. 
More practical timescale of the interaction will be larger than 
this when we consider the concerned physical and chemical processes. 
A possible preliminary guess on timescales is a crossing timescale 
of $\sim$ 10$^6$ yrs for each CO cloud. This is roughly consistent 
with that from the ratio of the velocity and spatial dispersions of 
the clouds along the jet axis although the estimates should be 
crude at best and much affected by the initial conditions in the 
width of the relativistic jet and fluctuations of 
the H\emissiontype{I} gas in velocity and density. For instance, 
the jet may expand in width at 100--200 pc from the engine, causing 
the increase of the molecular-jet width. In such a case the timescale 
should be smaller than the above.

Another possible constraint is the timescale for CO formation in 
the H\emissiontype{I} gas via interstellar shocks, which requires 
$\sim$ 10$^5$--10$^{6}$ yrs depending on density. Recent studies 
of high latitude molecular clouds associated with a shell created 
by stellar wind offer an observational support for molecular 
formation by the shock compression (Yamamoto et al. 2003, 2006). 
A theoretical study shows that the shock-produced CO clouds should 
have peak velocities and velocity dispersions similar to those of 
the ambient H\emissiontype{I} gas (see Koyama \& Inutsuka 2002). 
The observed velocity dispersion among the CO clouds, $\sim$ 2--5 
km s$^{-1}$, seems consistent with this. In order to obtain a more 
detailed understanding of the physical processes in the interaction, 
we definitely need to elaborate on the physical and chemical 
processes in the shock which is beyond the scope of this paper. 
Theoretical studies of magneto-hydrodynamics of such interaction 
have been made for protostellar jet (Shibata \& Uchida 1990) and 
are to be extended to the relativistic jet with appropriate 
modifications of physical parameters. Such studies will shed more 
light on the processes discussed above.

Another issue to be considered in SS433 is the timescale of the SNR. 
The lifetime of W50 is generally assumed to be an order of 10$^{4}$ 
yrs (Safi-Harb \& Ogelman 1997; Safi-Harb \& Petre 1999) but the 
present model suggests that the lifetime of the SS433 jet may be an 
order of magnitude larger than the assumed lifetime of the SNR. It 
is important to reinvestigate a possible range of the timescale of 
the SNR allowed under the present observed physical parameters.

\subsubsection{Energetics}

We shall examine the energy balance in the interaction by focusing on 
the SS433 southern clouds where the interaction is more clearly 
identified than in the north. In Table 3, the kinetic power of the 
southern SS433 jet is estimated to be $\sim$ 1$\times$10$^{46}$ 
erg yr$^{-1}$ for the speed and mass flow rate derived from the 
H$\alpha$ emission (Marshall et al. 2002). The energy deposit in the 
molecular gas through the interaction is roughly estimated to be 
$\sim$ 10$^{48}$ erg for an assumed expansion velocity of $\sim$ 5 
km s$^{-1}$, the velocity dispersion among the CO clouds (Table 1). 
This corresponds to 10$^{-3}$ of the total energy of the jet if 
its duration of is $\sim$ 10$^{5}$ yrs, suggesting that the energy 
requirement is well satisfied under the SS433 jet properties and that 
most of the deposited energy is radiated away. MJG348.5 also has 
similar expansion energy of the molecular clouds and is explicable 
with the parameters of the SS433 jet. If we take into account the 
atomic mass, this energy may become somewhat larger but not by more 
than an order of magnitude.

\subsubsection{The interaction with H\emissiontype{I}}

In the present scenario, the jet should lose its momentum through 
the interaction with the pre-existent H\emissiontype{I} gas. As long 
as the jet is running at a velocity close to the light speed, we 
expect the alignment of the formed molecular clouds along the axis 
because the velocity of the jet is much faster than the local 
turbulent motion of several km s$^{-1}$ in the undisturbed 
H\emissiontype{I} gas (see, e.g., Figures 2(b) and 6(a)). When 
the deceleration becomes significant so that the expanding speed 
becomes less than the local turbulent velocity of H\emissiontype{I}, 
the distribution of the forming clouds may become dominated by the 
local ambient velocity field. The large velocity dispersions are 
in fact seen only towards the tips of the jet, i.e., in SS433 south 
cloud (SS433-S6), MJG348.5-N and MJG348.5-S3, and the magnitude of 
the velocity dispersion, several km s$^{-1}$, is roughly consistent 
with the local turbulent motion of H\emissiontype{I}. These enhanced 
dispersions at the tips may indicate the deceleration at the end of 
the interaction. 

In SS433, it is somewhat unusual that the H\emissiontype{I} gas is 
distributed at such high galactic latitude of $\sim$ \timeform{-4D} 
prior to the interaction.  We suggest that this may be due to the 
stellar winds by an early type star which caused supernova explosion 
prior to the formation of the relativistic jet. Such a gas shell 
is found in CO, H\emissiontype{I} and the dust emission in Pegasus 
loop which is driven by an early B-type star (Yamamoto et al. 2006). 
The H\emissiontype{I} gas in the south is perhaps part of the shell 
created by the stellar wind of the supernova precursor.

The mass of each molecular cloud formed by the interaction is
an order of 10$^2$--10$^3$ $M_{\odot}$ and is comparable to
the mass of the HI gas in the cylinder of the length of 10 pc and radius of 10 pc which is typical size of molecular clouds having density of $\sim$ 10--30 cm$^{-3}$.

\subsubsection{Tilt of the jet}

It is suggested that the SS433 jet is tilted to the line of sight by 
12 degrees in the sense that the southern part is closer to us 
(Hjellming \& Johnston 1981). This is qualitatively consistent with 
that the southern clouds have smaller velocities than the northern 
clouds by $\sim$ 10 km s$^{-1}$, while the nominal difference 
$\sim$ 500 pc between the northern and southern clouds in the 
kinematic distance seems too large, where the radial velocity is 
perhaps affected by the motion other than the galactic rotation like 
the expansion of the H\emissiontype{I} gas noted above. We should in 
addition recall that there is a small difference in the projected 
angle between the molecular jet and the X-ray jet by 10 degrees. 
This may be ascribed to a long-term precession of the axis, or 
alternatively, to the relative motion of the ambient 
H\emissiontype{I} gas.

In MJG348.5, the northern part has blue-shifted velocities compared 
to the southern part by $\sim$ 15 km s$^{-1}$. This is explicable as 
that the jet axis is somewhat tilted to the line of sight in the 
sense that the northern part is closer to us, while we should be 
cautious about the large uncertainties in the kinematic distance 
again.

\subsubsection{The candidates for the driving source in MJG348.5}

In SS433, the driving engine is most likely SS433 itself located at 
(\timeform{39D.69}, \timeform{-2D.24}) in ($l$, $b$). On the other 
hand, such a compact object are not uniquely identified in MJG348.5. 
There are some SNRs and X-ray/$\gamma$-ray sources towards MJG348.5 
in the literature (see section 4.2).  We note that the distances to 
the three SNRs, CTB37A , CTB37B, and G348.5$-$0.0, are estimated by 
various methods to be $\sim$ 3.1 kpc, $\sim$ 4.8 kpc, and $\sim$ 
10 kpc, respectively (e.g., Caswell et al. 1975; 
Vermeulen et al. 1993; Reynoso \& Mangum 2000). These values are 
different from the present kinematic distance, 6 kpc, although there 
is a large uncertainty in the kinematic distance towards the center 
of the Galaxy.  It should be important to make detailed observations 
of the energetic sources in the field at radio, X-ray and $\gamma$-ray 
wavelengths. The present scenario suggests that the interaction 
occurred $\gtrsim$ 10$^5$ yrs ago, possibly longer than the timescale 
of the currently observed energetic objects mentioned above and it 
is also possible that the driving source of MJG348.5 is not 
active now.  Based on these considerations, we shall postpone 
discussing the driving engine until more details of the high energy 
objects are revealed towards the center of MJG348.5.

\subsection{Further Implications}

The most developed jet in the Galaxy known to date is that in SS433 
(e.g., Margon 1984; Margon \& Anderson 1989; Kotani 1998). GRS1915+105 
is another possible candidate if the association of the IRAS sources 
are correct (Chaty et al. 2001). Such an example showing an elongated 
jet of more than 10 pc in length is very rare except for SS433 and 
possibly GRS1915+105 in the Galaxy. Relativistic jets observed in the 
X-ray or radio synchrotron emission are usually much smaller in 
their size than the SS433 jet. These include pulsar wind nebulae 
(=PWN) like Crab and radio/X-ray jets including Cygnus X-1 
(Gallo et al. 2005) and are in the order of several pc at most.  
The length of the jet and physical properties of molecular clouds 
between SS443 and MJG348.5 are nearly the same and the physical 
parameters of the driving source of MJG348.5 may be similar to those 
of SS433. 

The present discoveries have shown the first candidates for the 
400-pc-scale jet from a stellar remnant, three to five times longer 
than previously known in the Galaxy.  It is also noteworthy that the 
relativistic jet may be able to interact with the interstellar medium 
to form molecular clouds. The present findings have opened a new 
possibility to use the millimeter CO emission to search for 
relativistic jet from a black hole or a neutron star. The time scale 
of the jet formation is $\sim$ 10$^{5-6}$ yrs, offering a probe for 
a past episode of relativistic jet.  At the moment we have only two 
cases of this kind of molecular jet. This is not surprising since 
the CO surveys with high angular resolutions has rarely been made for 
the latitude range above \timeform{1D} in $\mid$$b$$\mid$ extensively 
(Jackson et al. 2006).  High resolution CO surveys with a large 
latitude coverage has a potential to broaden our knowledge on black 
holes and neutron stars and their related activities in the Galaxy. 

We recall the other unusual feature seen in Figure 5 towards 
($l$, $b$) $\sim$ (\timeform{347D.7}, \timeform{1D.6}) ( section 4.1). 
This cloud may represent the third sample of this type; we shall note 
that there is a hint of an H\emissiontype{I} protrusion, a possible 
counterpart without CO, below the galactic plane at 
$b$ $\sim$ \timeform{-1D} at a similar longitude 
(see Figure 8, ($l$, $b$) = (\timeform{347D.5} -- \timeform{347D.7}, 
\timeform{-1D.5} -- \timeform{1D.5})). We cannot exclude a possibility 
that this cloud is an object similar to MJG348.5 although the 
supportive evidence is yet weaker than in the other two at the moment.
	
The process in the present scenario is analogous to what is familiar 
as the "vapor trail" produced by a jet airplane in the Earth's 
atmosphere although the detailed mechanism of formation is obviously 
different; the vapor trail is solid or liquid water condensed under 
the influence of the hot ejected matter by a jet engine. A common 
phenomenological property between the present jet and the vapor 
trail is that the remnant trail survives over a much longer time scale 
than that of the interaction, holding the position of the path of the 
interaction. We shall propose to call the present jet as "galactic 
vapor trail" because of such similarity.

\section{Conclusions}

Main conclusions of the present study are summarized as follows;

\begin{enumerate}
\item We have carried out a detailed analysis in two large areas of 
$\sim$ 25 square degrees around SS433 and of $\sim$ 18 square degrees 
towards $l$ $\sim$ \timeform{348D.5} using the NANTEN Galactic Plane 
Survey $^{12}$CO($J$=1--0) dataset.  We have discovered two groups of 
well aligned molecular clouds; ten molecular clouds along the X-ray 
jet axis of SS433 and four molecular clouds towards $l$ $\sim$ 
\timeform{348D.5} over $\sim$ 4 degrees in $b$ with $\sim$ 10 arcmin 
width perpendicular to the Galactic plane. 

\item We suggest that the aligned molecular clouds in the SS433 region 
represent the molecular gas created by the interaction between the 
relativistic jet and the interstellar H\emissiontype{I} gas. The 
present clouds are extended by $\sim$ 150 pc at a kinematic distance 
of 3 kpc, suggesting that the relativistic jet is more extended by a 
factor of about three towards the south than the X-ray jet. 

\item We apply the same relativistic-jet model to the MJG348.5 clouds 
in order to explain the alignment over $\sim$ 400 pc at a kinematic 
distance of $\sim$ 5 kpc and suggest that the clouds were created by 
the interaction between the hypothetical relativistic jet and the 
H\emissiontype{I} gas. Four high energy objects, three SNRs and a 
gamma ray source, in the galactic plane are considered as the 
candidate for the driving engine while none of them has known 
relativistic jet at the moment.

\item In either case, the estimated kinetic energy of the molecular 
clouds is $\sim$ 10$^{48}$ erg, corresponding to 1\% of the total 
kinetic energy released in relativistic jet from SS433 over 
$\sim$ 10$^{5}$ yrs. This suggests that the energy of the relativistic 
jet is large enough to form such molecular clouds.

\item We suggest that the present findings open a new possibility 
to search for candidates of a neutron star or a black hole over 
more than 10-times longer timescales than the direct detection of 
high energy radiation from these compact objects. We name the 
phenomenon as "galactic vapor trail" from the analogy with the 
vapor trail created by jet airplanes.

\end{enumerate}

\bigskip

We greatly appreciate the hospitality of all staff members of the Las
Campanas Observatory of the Carnegie Institution of Washington. The
NANTEN telescope is operated based on a mutual agreement between
Nagoya University and the Carnegie Institution of Washington. We also
acknowledge that the operation of NANTEN can be realized by
contributions from many Japanese public donators and companies. 
This work is financially supported in part by a Grant-in-Aid for Scientific Research from the Ministry of Education, Culture, Sports, Science and Technology of Japan (No.\ 15071203) and from JSPS (No.\ 14102003, core-to-core program 17004 and No.\ 18684003).

\appendix

\section*{Probability of the aligned clouds}

Here we shall give a quantitative estimate of the probability for an alignment of molecular clouds along a jet axis.

We shall first assume that a jet axis is defined by two clouds at lower latitude (e.g., MJG348.5) or by the driving engine and another cloud at lower latitude (e.g., SS433). We shall then estimate the probability to find another cloud at higher latitude on the jet axis including the agreement in velocity.

In case of MJG348.5, the jet axis is defined by MJG348.5-S1 and MJG348.5-S2, and MJG348.5-S3 "happens" to be located on the axis at higher latitude above 2 degrees.  Figure 7 does illustrate that such a CO cloud like MJG348.5-S3 at above 2 degrees is only one in the field presented, very rare on both the negative and positive latitudes. The only one exception is seen at ($l$, $b$) $\sim$ (\timeform{356D}, \timeform{2D}) which is the top of the proposed magnetic floatation loop further away in the galactic center (Fukui et al. 2006). We estimate that only one cloud, MJG348.5-S3, is found for an area of 500 pc $\times$ 500 pc projected on the galactic plane (Figure 10(a)), where 500 pc corresponds to a half of the projected length along the longitude as well as along the line of sight estimated from the velocity span covered, from $-$100 km s$^{-1}$ to $-$70 km s$^{-1}$, shown in Figure 7. 

By adopting a typical cloud size 10 pc as a size of a 2 dimensional cell, we shall estimate the probability to find S3 on the jet axis as to be 10pc/500pc $\times$ 10pc/500pc = 1/50 $\times$ 1/50 $\sim$ 4$\times$10$^{-4}$. This is actually a secure upper limit since there is another cloud N on the positive latitude, whose alignment makes the probability of MJG348.5 even smaller.  

In case of SS433, the jet axis is defined by two objects, SS433 itself and, for simplicity, a set of clouds SS433-S1--SS433-S4 at $b$ below $-$4 degrees which has a typical spatial dispersion of 10 pc around the jet axis (Figure 10(b)).  Another set of clouds, SS433-S5 and SS433-S6, then "happens" to be located on the axis at $b$ above $-$4 degrees.  Figure 3 does illustrate that such a set of CO clouds like SS433-S5 and SS433-S6 at above $-$4 degrees is only one in the field presented. We again estimate that a set of CO clouds is found for an area of 300 pc $\times$ 300 pc projected on the galactic plane (Figure 10(b)), where 300 pc corresponds to a half of the along longitude projected on the sky as well as the depth along the line of sight estimated from the velocity span covered, from 40 km s$^{-1}$ to 60 km s$^{-1}$, shown in Figure 3. By using the same argument as above, we estimate the probability to find the clouds on the jet axis to be 10pc/300pc $\times$ 10pc/300pc= 1/30 $\times$1/30 $\sim$ 10$^{-3}$.

To summarize, we argue that the probability of the present aligned clouds is as small as 10$^{-3}$ -- 3$\times$10$^{-4}$.

\clearpage

\begin{figure}
  \begin{center}
     \FigureFile(80mm, 80mm){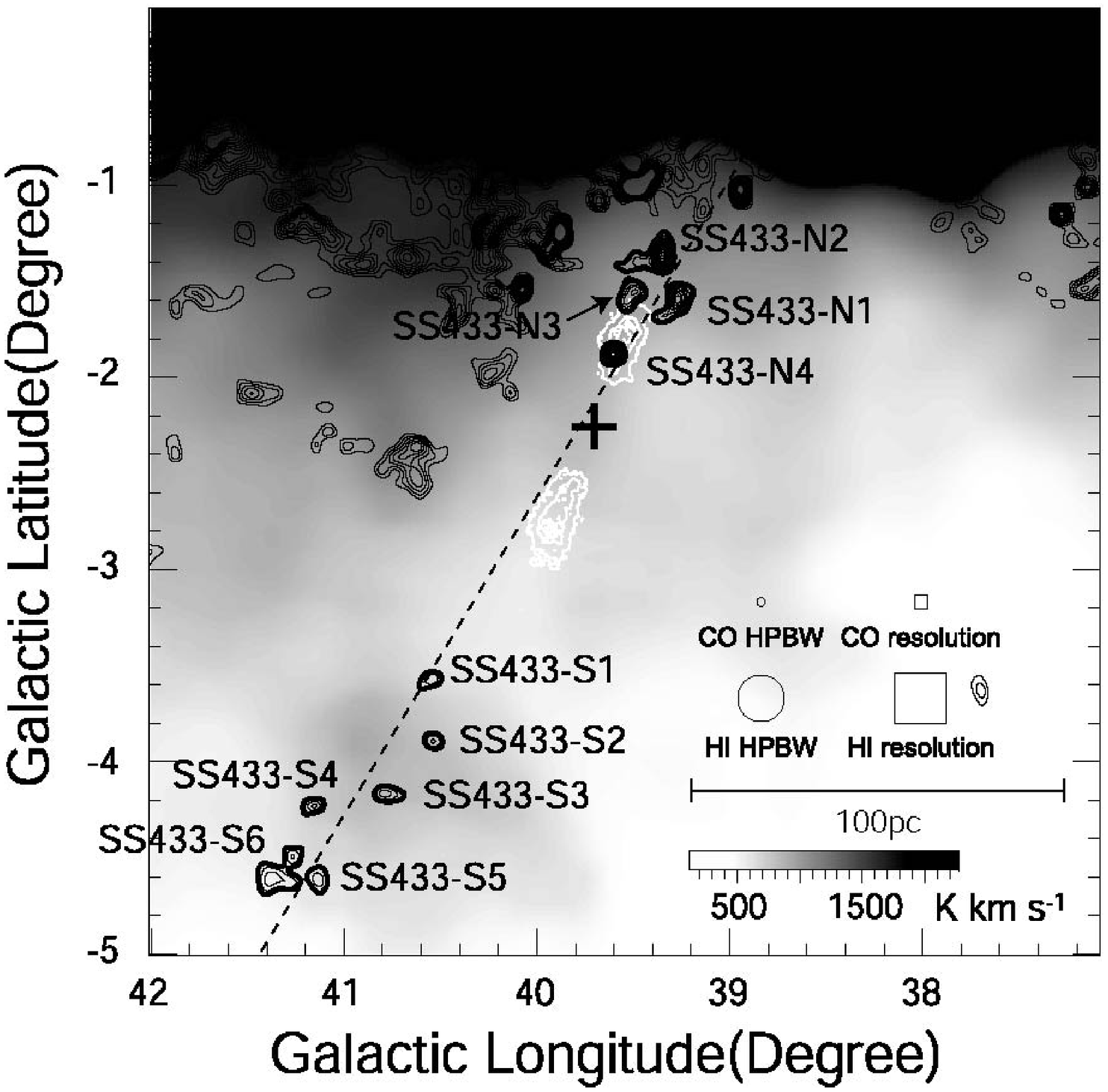}
  \end{center}
  \caption{The integrated intensity map of $^{12}$CO($J$=1--0) 
(black contours) superposed on the H\emissiontype{I} (gray scale) 
whose velocity range is 40 to 60 km s$^{-1}$.
Boundaries of SS433-N1 -- SS433-N4, and SS433-S1 -- SS433-S6 are 
shown in thick contour lines.
The contours of the CO are illustrated every 1.8 K km s$^{-1}$  
from 3.6 K km s$^{-1}$.
The black cross indicates the position of SS433, and white contours 
indicate ASCA X-ray image of the SS433 lobes. 
The dashed line indicates the result of the linear regression fit to 
the north and south clouds.}
\end{figure}

\begin{figure}
  \begin{center}
     \FigureFile(80mm, 100mm){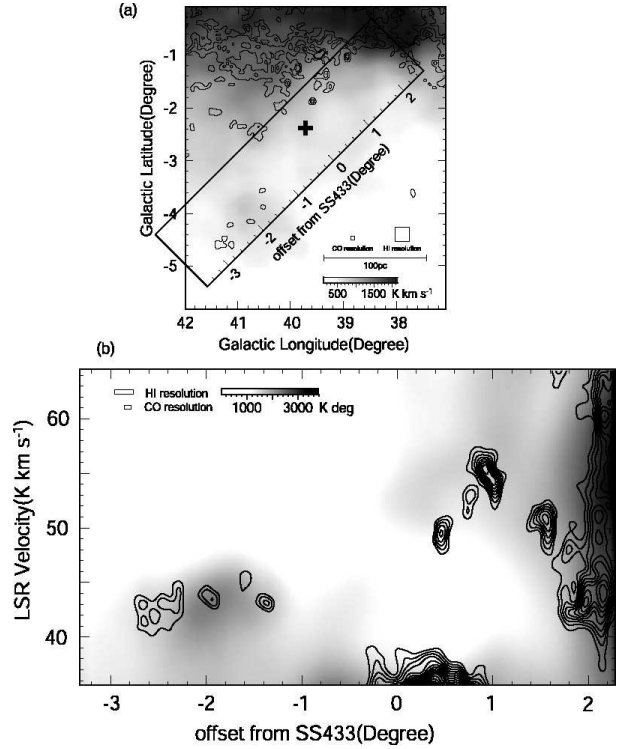}
  \end{center}
  \caption{(a)The same as Figure 1, but contour levels of CO are 
every 10 K km s$^{-1}$ from 3.6 K km s$^{-1}$. 
(b)The position-velocity diagram of $^{12}$CO($J$=1--0) 
emission (contours) superposed on the H\emissiontype{I} (gray scale)
which is integrated in direction of the long side of rectangle in (a).
The contours of the CO are illustrated every 1.5 K deg from 5.5 K deg.
Offsets are relative to the position of SS433.}
\end{figure}

\begin{figure}
  \begin{center}
     \FigureFile(160mm, 90mm){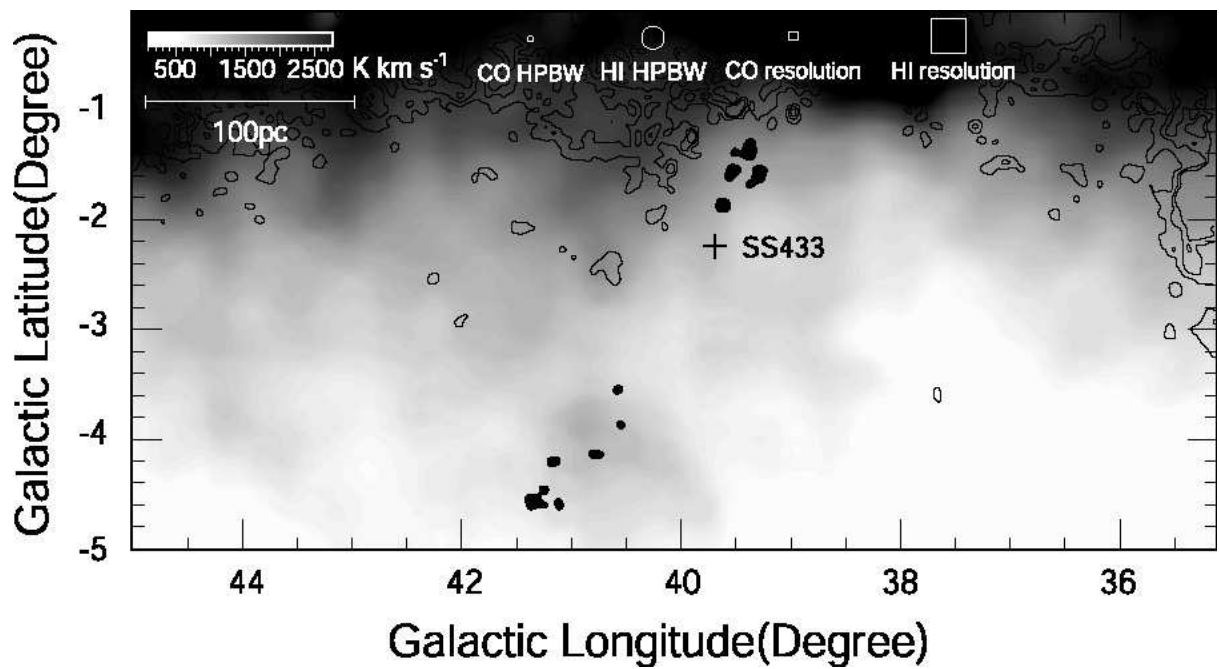}
  \end{center}
  \caption{The large scale integrated intensity map of 
$^{12}$CO($J$=1--0) (black contours) superposed on the 
H\emissiontype{I} (gray scale) whose velocity range is 
40 to 60 km s$^{-1}$. The contours of the CO are illustrated every 10 
K km s$^{-1}$ from 5.8 K km s$^{-1}$.
The black cross indicates the position of SS433.
North and south clouds are filled in black.}
\end{figure}

\begin{figure}
  \begin{center}
     \FigureFile(80mm, 80mm){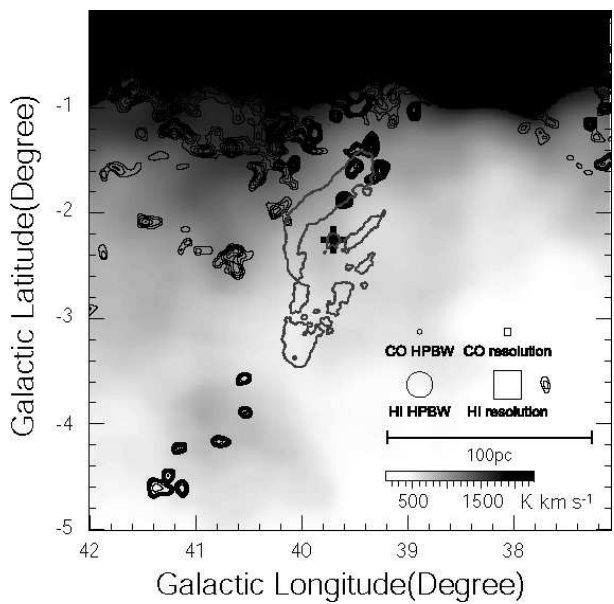}
  \end{center}
  \caption{The same as Figure 1, but VLA radio continuum image 
of W50 at 4850MHz is superposed instead of ASCA X-ray image.}
\end{figure}

\begin{figure}
  \begin{center}
     \FigureFile(80mm, 160mm){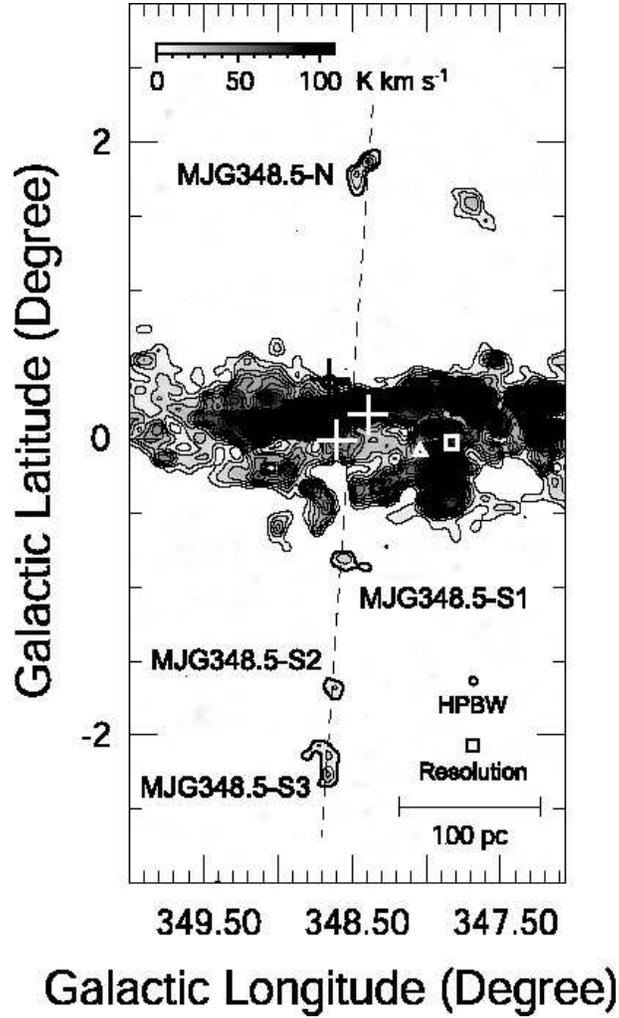}
  \end{center}
  \caption{The integrated intensity map of $^{12}$CO($J$=1--0) 
whose velocity range is $-$100 to $-$70 km s$^{-1}$. The contours 
of CO are illustrated every 4.2 K km s$^{-1}$ from 4.2 
K km s$^{-1}$ (5$\sigma$). Four molecular clouds, MJG348.5-N, 
MJG348.5-S1, MJG348.5-S2, and MJG348.5-S3, are shown in the 
figure and boundaries of them are shown in thick contours. 
The crosses indicate the positions of the supernova remnants, 
CTB37A: ($l$, $b$) $\sim$ (\timeform{348D.39}, \timeform{0D.16}), 
CTB37B: ($l$, $b$) $\sim$ (\timeform{348D.65}, \timeform{0D.40}), 
and G348.5$-$0.0: ($l$, $b$) $\sim$ (\timeform{348D.60}, 
\timeform{-0D.01}).  The circle and triangle indicate the positions 
of the $\gamma$-ray sources identified by H.E.S.S. and EGRET, 
respectively. The squares indicate the positions of X-ray sources 
identified by ROSAT. These symbols are illustrated by white and 
black in the region where the integrated intensity of CO is high 
and low, respectively. These source are listed in Table 2.}
\end{figure}

\begin{figure}
  \begin{center}
     \FigureFile(80mm, 50mm){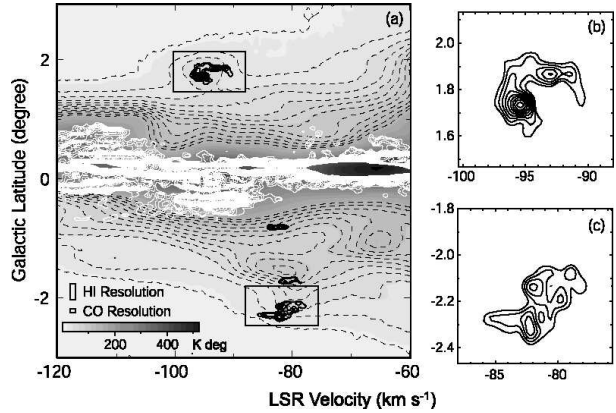}
  \end{center}
  \caption{(a) The velocity-latitude diagram  of $^{12}$CO($J$=1--0) 
emission (black and white contours) superposed on 
the H\emissiontype{I} (gray scale and dotted contours) whose 
integrated range in the Galactic longitude is from \timeform{348D.267} 
to \timeform{348D.800}.  MJG348.5-N, MJG348.5-S1, MJG348.5-S2, and 
MJG348.5-S3 are shown in black solid contours.  The contour levels of 
CO are shown up to 20 K deg every 0.6 K deg from 1.5 K deg  and those 
of the H\emissiontype{I} up to 145.2 K deg every 12 K deg from 13.2 
K deg. The resolution of CO and H\emissiontype{I} are shown in the 
figure.  (b), (c)  Close-up views of the upper and lower boxes in (a) 
for only $^{12}$CO. The contour levels of $^{12}$CO in (b) and (c) 
are the same as (a).}
\end{figure}

\begin{figure}
  \begin{center}
     \FigureFile(160mm, 90mm){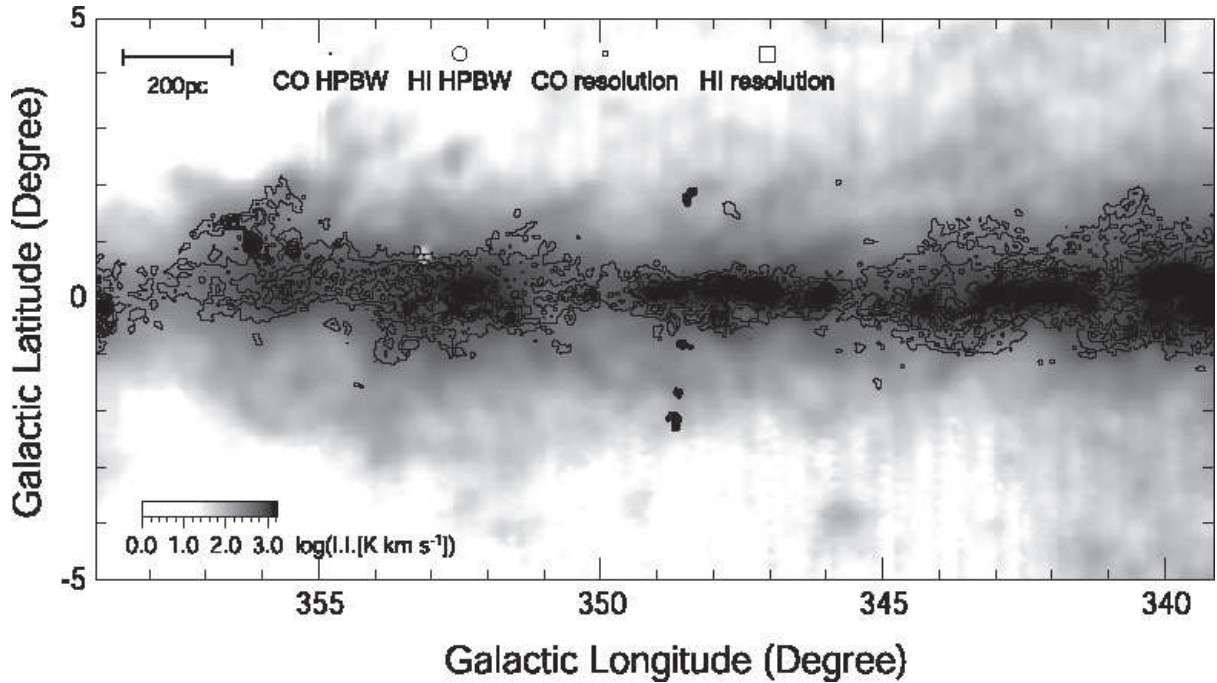}
  \end{center}
  \caption{The large scale integrated intensity map of $^{12}$CO 
($J$=1--0) (solid contours) superposed on H\emissiontype{I} (dashed 
contours and gray scale) whose velocity range is $-$100 to 
$-$70 km s$^{-1}$. The intensity of the CO is illustrated in linear 
scale and that of the H\emissiontype{I} is illustrated in logarithmic 
scale. The four molecular clouds, MJG348.5-N, MJG348.5-S1, MJG348.5-S2 
and MJG348.5-S3 toward $l$ $\sim$ \timeform{348D.5} are filled in 
black.}
\end{figure}

\begin{figure}
  \begin{center}
     \FigureFile(80mm, 80mm){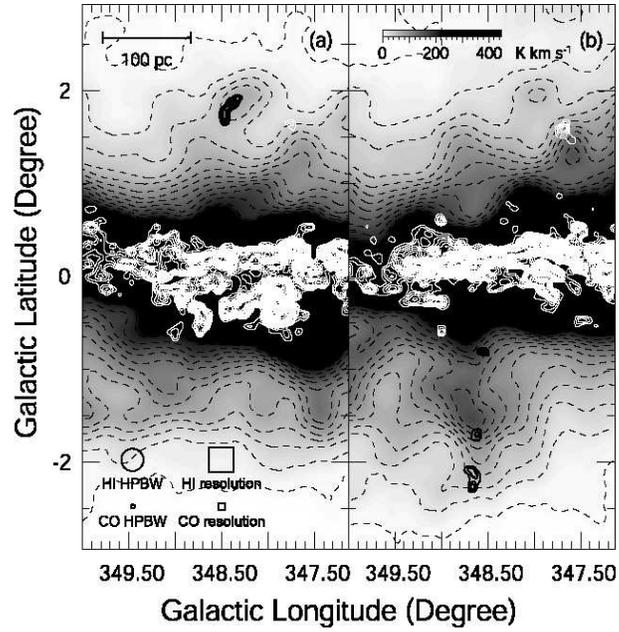}
  \end{center}
  \caption{The integrated intensity channel maps of 
$^{12}$CO($J$=1--0) (black and white contours) superposed on 
that of H\emissiontype{I} (gray scale and black dashed contours). 
MJG348.5-N, MJG348.5-S1, MJG348.5-S2 and MJG348.5-S3 are shown in 
black solid contours.  The velocity ranges of 
both CO and H\emissiontype{I} data in (a) and (b) are $-$100 to 
$-$85 km s$^{-1}$ and $-$85 to $-$70 km s$^{-1}$, respectively. 
The contours of CO and H\emissiontype{I} are illustrated every 
2.1 K km s$^{-1}$ from 4.2 K km s$^{-1}$ and every 
15 K km s$^{-1}$ from 15 K km s$^{-1}$, respectively. }
\end{figure}

\begin{figure}
  \begin{center}
     \FigureFile(160mm, 160mm){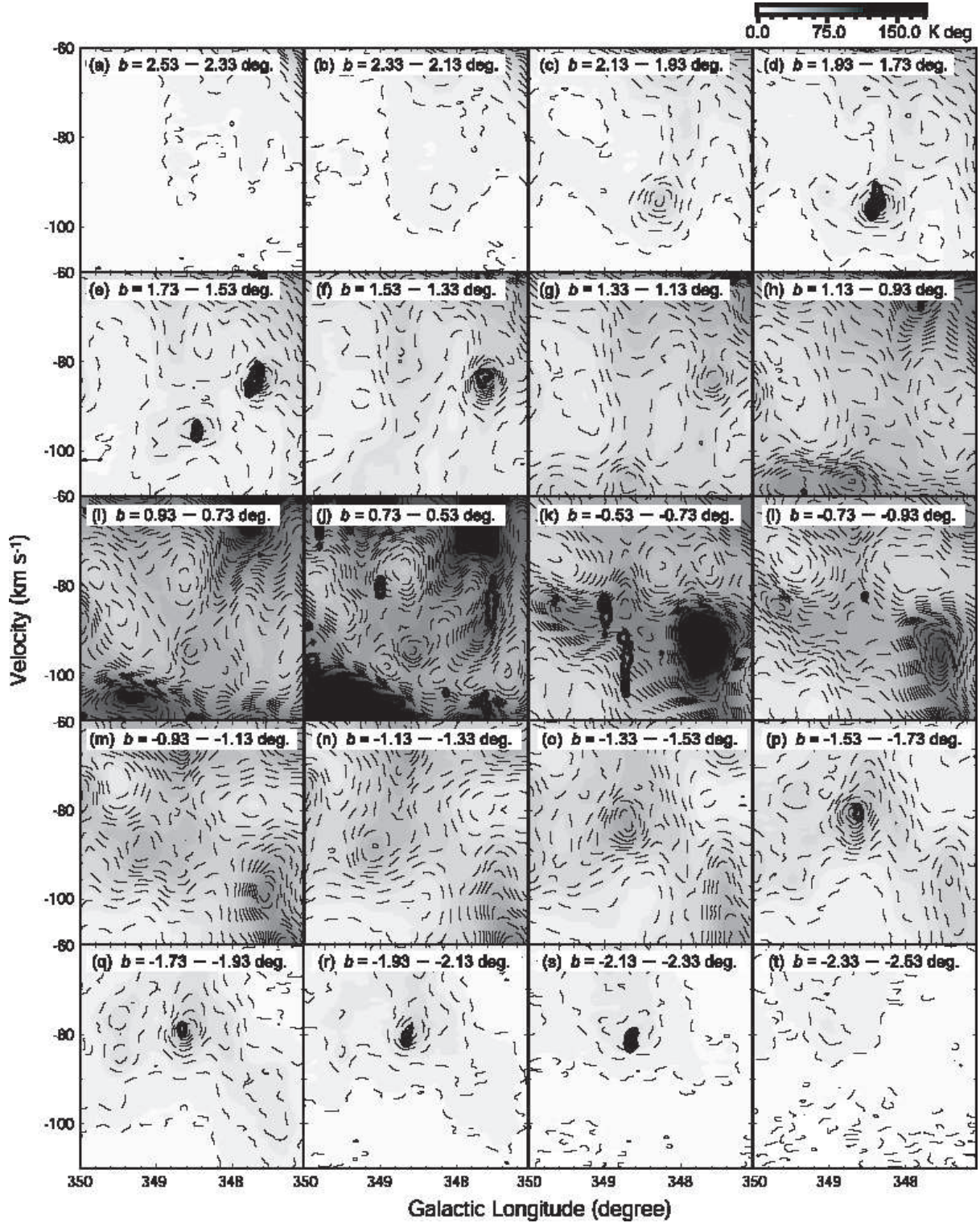}
  \end{center}
  \caption{Channel maps of logitude--velocity diagram from 
2.53 to $-$2.53 degree except within $\pm$0.53 degree every 
0.2 degree in Galactic latitude.  The velocity of H\emissiontype{I} 
and CO is smoothed at a resolution of 1 km s$^{-1}$. The range of 
the galactic latitude is shown in the upper side of each channel map. 
Gray scale and dashed contours are H\emissiontype{I} and solid 
contours are CO. MJG348.5-N, S1, S2, and S3 appear to the channel 
maps of (d) and (e), (l), (p) and (q), and (r) and (s), respectively. 
The contour levels of CO and H\emissiontype{I} are illustrated every 
1.5 K degree from 2.0 K degree and every 2.0 K degree every 4.0 K 
degree, respectively.}
\end{figure}

\begin{figure}
  \begin{center}
     \FigureFile(160mm, 80mm){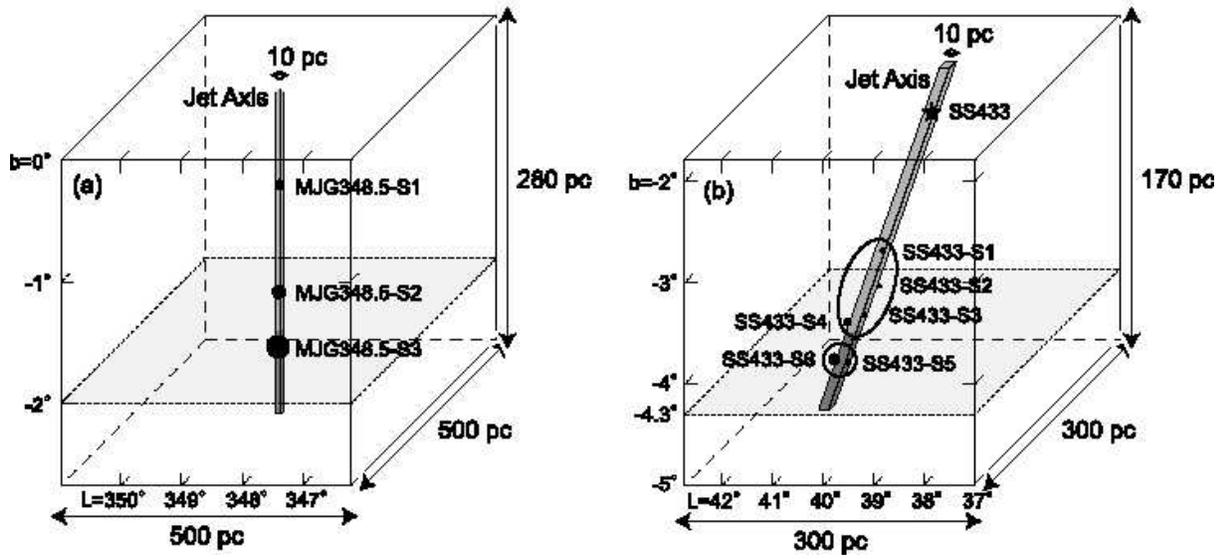}
  \end{center}
  \caption{Schematic view of the space in case of (a) MJG348.5 and (b) SS433. Black filled circles indicate
each molecular cloud identified. The planes of (a)  $b$ = \timeform{-2D} and (b) $b$ = \timeform{-4D.3} are illustrated by 
light gray.
Gray prisms indicate the direction of the jet. Lower part of the each gray prism in darker gray
shows the part lower than (a) $b$ = \timeform{-2D} and (b) $b$ = \timeform{-4D.3}. }
\end{figure}

\clearpage

\catcode`#=\active \def#{\phantom{0}}
\begin{table*}
\begin{center}
\caption{Physical Properties of $^{12}$CO Clouds}
\begin{tabular}{ccccccccccc}
\hline
No. & $l$ & $b$ & $T_{\rm R}^{\ast}$ & $\Delta$V & $V_{\rm LSR}$ & 
$N$(H$_2$) & $R$\footnotemark[$*$] & $M_{\rm CO}$\footnotemark[$*$] & 
$M_{\rm vir}$\footnotemark[$*$] & $t_{\rm cross}$\footnotemark[$*$] \\
& (\timeform{D}) & (\timeform{D}) & (K) & (km s$^{-1}$) & 
(km s$^{-1}$) & (10$^{21}$ cm$^{-2}$) & (pc) & ($M_{\odot}$) & 
($M_{\odot}$) & (Myr) \\
\hline
SS433-N1 & 39.27 & $-$1.60 & #4.0 & 3.1 & #55.8 & 3.9 & #5.1 & 
#2500 & #6000 & 2.3\\
SS433-N2 & 39.33 & $-$1.33 & #6.0 & 2.9 & #53.7 & 3.0 & #5.5 & 
#3100 & 14000 & 1.6\\
SS433-N3 & 39.47 & $-$1.53 & #3.3 & 1.9 & #53.0 & 1.5 & #3.6 & 
#1300 & #5400 & 1.3\\
SS433-N4 & 39.60 & $-$1.87 & 10.0 & 2.3 & #49.4 & 4.8 & #3.6 & 
#2100 & #5000 & 1.4\\
SS433-S1 & 40.53 & $-$3.53 & #3.9 & 1.4 & #42.9 & 1.4 & #2.8 & 
##600 & #1200 & 1.9\\
SS433-S2 & 40.53 & $-$3.87 & #2.7 & 2.2 & #45.4 & 1.4 & #2.0 & 
##400 & #6400 & 0.5\\
SS433-S3 & 40.80 & $-$4.13 & #2.9 & 2.4 & #44.1 & 1.7 & #2.8 & 
##900 & #3300 & 1.2\\
SS433-S4 & 41.13 & $-$4.20 & #1.8 & 3.8 & #43.2 & 1.5 & #3.6 & 
#1200 & 21000 & 0.7\\
SS433-S5 & 41.13 & $-$4.60 & #1.2 & 4.2 & #44.8 & 1.1 & #2.8 & 
##500 & 16000 & 0.5\\
SS433-S6 & 41.33 & $-$4.53 & #2.4 & 2.6 & #42.1 & 1.4 & #5.6 & 
#2300 & 16000 & 1.6\\
MJG348.5-#N & 348.47 & #1.73 & #3.3 & 5.4 & $-$95.0 & #2.3 & 
12.5 & 14000 & 77000 & 2.3\\
MJG348.5-S1 & 348.53 & $-$0.80 & #1.9 & 3.4 & $-$82.4 & #1.6 & 
#3.9 & #1500 & #9700 & 1.2\\
MJG348.5-S2 & 348.60 & $-$1.67 & #1.6 & 3.3 & $-$80.0 & #1.1& 
#5.6 & #2000 & 12000 & 1.7\\
MJG348.5-S3 & 348.67 & $-$2.13 & #3.2 & 4.3 & $-$81.3 & #1.3 & 
#9.6 & #8300 & 26000 & 2.2\\
\hline
\multicolumn{11}{l}{\hbox to 0pt{\parbox{160mm}{\footnotesize
Col. (1) : Cloud number, Col. (2)--(3) : Cloud
peak ($l$, $b$) position, Col. (4) : Peak temperature, Col. (5) : Line
width of the composite spectrum, Col. (6) : Peak velocity of the
composite spectrum, Col. (7) : Column density of peak position, 
Col. (8) : Radius of the molecular cloud, 
Col. (9) : Mass of the molecular cloud, Col. (10) : 
Virial mass of the molecular cloud, Col. (11) : Crossing
time of the molecular cloud. Col. (4) to (6) are derived by 
using a single Gaussiun fitting. \\
\footnotemark[$*$] The distance is assumed as 3 kpc on SS433 and 6 kpc 
on MJG348.5. In details, see Sec. 3.2 and 4.2.
}\hss}}
   \end{tabular}
  \end{center}
\end{table*}

\clearpage

\begin{table*}

\begin{center}
\rotatebox{90}{\begin{minipage}{\textheight}
\caption{List of Supernova Remmants and High Energy Sources toward 
MJG348.5 in the Galactic Plane\footnotemark[$*$]}
\begin{tabular}{cccccccc}
\hline
Name & \multicolumn{4}{c}{Positions} & Type of Source & 
Wavelength of detection & Associated Objects \\
\cline{2-5} \\
& $l$ & $b$ & $\alpha(J2000)$ & $\delta(J2000)$ & & & \\
\hline
G348.5+0.1(CTB37A)\footnotemark[$\dagger$] & \timeform{348D.39} & 
\timeform{0D.16} & \timeform{17h14m6s.0} & \timeform{-38D32'0''} & 
SNR & Radio-Conti. (90cm) & \\
G348.5-0.0\footnotemark[$\dagger$] & \timeform{348D.60} & 
\timeform{-0D.01} & \timeform{17h15m26s.0} & \timeform{-38D28'2''} & 
SNR & Radio-Conti. (90cm) &  \\
G348.7+0.3 (CTB37B)\footnotemark[$\dagger$] & \timeform{348D.65} & 
\timeform{0D.40} & \timeform{17h13m55s.1} & \timeform{-38D10'59''} & 
SNR & Radio-Conti. (90cm) & \\
3EG J1714-3857\footnotemark[$\ddagger$] & \timeform{348D.04} & 
\timeform{-0D.09} & \timeform{1714m5s.4} & \timeform{-38D57'54''} & 
$\gamma$-ray & $\gamma$-ray ($E$ $>$ 100 MeV) \\
H.E.S.S. J1713-381\footnotemark[$\S$] & \timeform{348D.65} & 
\timeform{0D.38} & \timeform{17h13m58s.1} & \timeform{-38D11'43''} & 
$\gamma$-ray & $\gamma$-ray ($E$ $>$ 100 GeV)  & CTB37A, CTB37B, 
G348.5-0.0 \\
1RXS J171312.8-390553\footnotemark[$\|$] & \timeform{347D.83} & 
\timeform{-0D.03} & \timeform{17h13m12s.8} & \timeform{-39D5'54''} & 
X-ray &   X-ray & \\
1RXS J171551.8-385843\footnotemark[$\|$] & \timeform{348D.23} & 
\timeform{-0D.38} & \timeform{17h15m51s.9} & \timeform{-38D58'43''} & 
X-ray & X-ray & \\
1RXS J171557.7-385152\footnotemark[$\|$] & \timeform{348D.33} & 
\timeform{-0D.33} & \timeform{17h15m57s.7} & \timeform{-38D51'51''} & 
X-ray &   X-ray & \\
1RXS J171354.4-381740\footnotemark[$\|$] & \timeform{348D.56} & 
\timeform{0D.33} & \timeform{17h13m54s.4} & \timeform{-38D17'38''} & 
X-ray &  X-ray & CTB37B \\
\hline
\multicolumn{8}{l}{\hbox to 0pt{\parbox{160mm}{\footnotesize
\footnotemark[$*$] Sources within \timeform{1D} from ($l$, $b$) 
$\sim$ (\timeform{348D.5}, \timeform{0D.0}) are listed \\
\footnotemark[$\dagger$] Kassim, Baum, \& Weiler (1991) \\
\footnotemark[$\ddagger$] Hartman et al. (1999) \\
\footnotemark[$\S$] Aharonian et al. (2006) \\
\footnotemark[$\|$] Voges et al. (1999)
}\hss}}
\end{tabular}
\end{minipage}}
\end{center}
\end{table*}

\clearpage

\begin{table}
\begin{center}
\caption{Physical Parameters of SS433}
\begin{tabular}{ccccc}
\hline
Velocity\footnotemark[$*$] & Mass flow rate\footnotemark[$\dagger$] & 
Momentum\footnotemark[$\ddagger$] & 
Kinetic power\footnotemark[$\dagger$] & Timescale\footnotemark[$\S$] \\
 & ($M_{\odot}$ yr$^{-1}$) & ($M_{\odot}$ km s$^{-1}$ yr$^{-1}$) & 
(erg yr$^{-1}$) & (yrs) \\
\hline
0.26$c$ & 1.5$\times$10$^{-7}$ & 1.2$\times$10$^{-2}$ & 
1.1$\times$10$^{46}$ & 2$\times$10$^{4}$ \\
\hline
\multicolumn{5}{l}{\hbox to 0pt{\parbox{80mm}{\footnotesize
\footnotemark[$*$] Margon \& Anderson (1989) \\
\footnotemark[$\dagger$] Marshall et al. (2002) \\
\footnotemark[$\ddagger$] (Velocity)$\times$(Mass flow rate) \\
\footnotemark[$\S$] Zealey et al. (1980)
}\hss}}
\end{tabular}
\end{center}
\end{table}

\end{document}